\documentclass[11pt]{article}
\usepackage{latexsym} 
\usepackage{epsfig}
\usepackage{a4}
\usepackage{amsfonts}
\makeatletter

\expandafter\let\expandafter\reset@font\csname reset@font\endcsname

\makeatother

\newcommand{\p}{^{\prime}}
\def\be{\begin{equation}}
\def\ee{\end{equation}}
\def\bea{\begin{eqnarray}}
\def\eea{\end{eqnarray}}

\def\D{\partial}
\def\'{\prime}

\def\ba{\begin{array}}
\def\ea{\end{array}}

\makeatletter
\def\binrel@#1{\begingroup
  \setboxz@h{\thinmuskip0mu
    \medmuskip\m@ne mu\thickmuskip\@ne mu
    \setbox\tw@\hbox{$#1\m@th$}\kern-\wd\tw@
    ${}#1{}\m@th$}%
  \edef\@tempa{\endgroup\let\noexpand\binrel@@
    \ifdim\wdz@<\z@ \mathbin
    \else\ifdim\wdz@>\z@ \mathrel
    \else \relax\fi\fi}%
  \@tempa
}
\let\binrel@@\relax
\def\overset#1#2{\binrel@{#2}%
  \binrel@@{\mathop{\kern\z@#2}\limits^{#1}}}
\def\underset#1#2{\binrel@{#2}%
  \binrel@@{\mathop{\kern\z@#2}\limits_{#1}}}
\newfont{\bbd}{msbm10 scaled\magstep1}
\begin{document}
%\vspace*{-2cm}

\begin{flushright}
LU-ITP 2004/12
\end{flushright}

\begin{center}
{\LARGE \bf  Parton interaction \\ in super Yang Mills theory
}

\vspace{1cm}
R. Kirschner
 \vspace{1cm}

 Institut f\"ur Theoretische Physik, Universit\"at Leipzig,
\\
Augustusplatz 10, D-04109 Leipzig, Germany

\end{center}

\vspace{1cm}
\begin{center}
{\bf Abstract}
\end{center}
\noindent  
We apply the effective action scheme to the leading parton
interactions in ${\cal N } = 1$  supersymmetric gauge theory.
The effective interaction in the Bjorken asymptotics at one loop is
written
in terms of parton superfield vertices explicitely symmetric with
respect to superconformal transformations.

%\vspace*{\fill}
%\eject
\vspace{1cm}

\section{Introduction}
\setcounter{equation}{0}

High-energy processes related to the Bjorken asymptotics \cite{DGLAP,ERBL}
are playing a major
role in the investigation of hadronic structure and interaction.

Aspects of conformal symmetry played an important role in the development of
the first concepts about the Bjorken limit and turned into an useful tool,
e.g. for finding multiplicatively renormalized operators \cite{Makeenko}
and for relating the forward to the non-forward evolution \cite{BuFKL,Muller}.
In combination with supersymmetry it allows to derive intersting relations
between the QCD DGLAP/ERBL evolution kernels \cite{BuFKL}. A recent review is
given in \cite{BKM}.

Scattering amplitudes   in the Bjorken asymptotics and the
renormalization of
composite operators contributing to this asymptotic expansion have been
studied in QCD  and in many field theory models in particular for
developing the
methods of   calculation and for investigating the symmetry properties
in the
asymptotics in comparison with the symmetries of the underlying theory.

A particular feature of symmetry encountered in some cases and
approximations is integrability, which attracted much interest in the
last years. It has been first noticed by Lipatov \cite{LevPadua} in studying the
perturbative Regge asymptotics of QCD \cite{BFKL} and formulated in
further detail by Faddeev and Korchemsky \cite{FK}. The methods of integrable
systems have been developed for the needs of such applications (see e.g.
\cite{DeVega:2001pu,Derkachov:2001yn})  and applied in particular to 
the renormalization of higher twist
composite operators, e.g. \cite{Braun:1998id,Belitsky:1999qh}. 
Questions related to the AdS/CFT
hypothesis motivated a series of studies of special composite operators
in ${\cal N }= 4$ super Yang- Mills theory, where also integrable
structures have been encountered \cite{Beisert:2003yb,Dolan:2003uh}.

There is a remarkable relation between the Regge and the Bjorken asymptotics
in  ${\cal N }= 4$ super Yang-Mills theory: The eigenvalues of the
Bjorken
evolution kernels, the twist-2 anomalous dimensions, can be obtained from the
eigenvalues of the Regge BFKL kernel by continuation to particular values of
the conformal weight \cite{Kotikov:2000pm}.

These developments motivated recent papers on the Bjorken asymptotics
in theories with supersymmetry 
\cite{Onishchenko:2003kh,Belitsky:2003sh,Belitsky:2004yg}
and provide also a motivation for the
present study.

In a recent paper \cite{Belitsky:2004yg}
Belitsky et al. have treated the supersymmetric ${\cal
N} = 1, 2, 4$ supersymmetric Yang- Mills theories emphasizing the
maximally extended ${\cal N} = 4$ case and the relation to
integrability. Their result on the two-parton interaction is
restricted to the term having
non-vanishing contribution in the configuration of parallel helicities,
which is the only one remaining in the ${\cal N} = 4$ case.

In a previous study \cite{DKBj} we have considered the Bjorken asymptotics
of QCD in the effective action approach with the aim of a close
comparison to the Regge asymptotics. The effective parton interaction
at one loop has been formulated in terms of vertices involving parton
fields residing on the one-dimensional light ray and kernels being
conformal 4-point functions. In this way we have achieved a formulation
of the one-loop DGLAP/ERBL evolution where conformal symmetry is
explicit.

The method of calculation and the way of symmetric formulation of the
result can be applied to other models. In this paper we consider
${\cal N} = 1 $ supersymmetric Yang-Mills theory (SYM).

In section 2 we summarize some results on QCD of the previous paper
\cite {DKBj}.  We discuss in some detail the appropriate formulation
of conformal symmetry. We present the main steps leading to the
symmetric form of the effective interaction in a way that allows to
demonstrate the analogy to the case of super Yang-Mills in the
following.

We avoid to compose the result for the super Yang-Mills theory out of
the ones for the gluon and fermion components. Instead we perform the
calculation in a simple super field language in section 3. In the first
step we obtain the result symmetric with respect to super translations.

In section 4 we write the generic form of the kernels of super conformal
symmetric operators. We give a basis of parton wave functions and of
composite parton field operators, related by a symmetric inner product,
corresponding to irreducible representations of the superconformal
symmetry. In this way we get the  eigenstates or multiplicatively
renormalized operators of the parton evolution operators. The
corresponding eigenvalues are the anomalous dimensions.

In section 5 we derive the super conformal symmetric interaction
vertices. This relies on the known generic form of symmetric kernels and
on the analogy to the gluodynamic case.

\section{Symmetry of parton interactions in QCD}
\setcounter{equation}{0}
\subsection{Calculation of kernels}

The kernels of the DGLAP/ERBL evolution  can be considered
as effective
vertices of parton interaction $1\p 2\p \rightarrow 1 2 $. The partons
carry merely one light-cone momentum component and can be represented by
fields living on the light-ray.   It is well known that this interaction
at one loop level is symmetric with respect to conformal
transformations acting on this light ray by M\"obius substitutions
(for a recent review see \cite{BKM}). The kernels have been identified
as conformal 4-point functions \cite{DKBj}. The parton fields on the
light ray emerge as particular modes of the QCD gluon and quark fields
by eliminating the redundant field components in light-cone gauge and by
separating the high virtual modes $A_p$ with differences of
coordinates  close to the light ray,

\be
\label{Ap}
 A(x_1) A(x_2)  \ \rightarrow A_{p1} (z_1) A_{p2} (z_2)  \ \delta_{Q}
(x_{12}^{\perp} ) \ \delta_{Q} (x_{12 +})
\ee
and the low virtual modes $A_{p\p}$ almost constant in a broad range
vertical to the light ray,
\be
A(x) \rightarrow A_{p\p} (z) \Delta_{m} (x^{\perp})
\ee
Here $ \delta_{Q} $ stands for narrow distribution of width $\sim Q^{-1}$
and $\Delta_{m}$ for a flat distribution of width $\sim m^{-1}$, $m \ll Q$.

We represent 4-vectors $x^{\mu} $ by their light cone components
$x_{\pm} $ and a complex number involving the transverse
components $ x_{\perp} = x_1 + i x_2$.
 In the case of the gradient vector
we change the notation in such a way to have $\partial_+ x_- =
\partial_- x_+ = 1, \partial_{\perp} x = \partial_{\perp}^* x^* = 1$.
 We choose the frame where the light-like
vector $q^{\prime}$ has the only non-vanishing component
$q^{\prime}_- = \sqrt {s/2}$ and $p^{\prime }$ the only non-vanishing
component $p_+^{\prime } = \sqrt {s/2}$.

The fields $A(x)$ denote in the case of gluons the transverse components
$A, A^*$  of the
vector potential in the light cone gauge $A_-=0$, where $A_+$ has been
integrated out. In the case of fermions they denote the light cone
components $f, f^*, \tilde f,  \tilde f^*$ defined as follows.

\bea
\psi = \psi_- + \psi_+, \ \ \gamma_- \psi_+ = \gamma_+ \psi_- = 0 \cr
\psi_+ = f u_{+-} + \tilde f u_{++}, \ \ \ \gamma^{\mu } = {1 \over
\sqrt s}
(\gamma_- q^{\prime \mu } + \gamma_+ p^{\prime \mu } ) + \gamma^{\mu }_{\perp }.
\eea

$u_{a,b}, \ \ a,b = \pm, $ is a basis of Majorana spinors,
\be
\gamma_+ u_{-,b} = \gamma_- u_{+,b} = 0,  \ \
\gamma u_{a, -} = \gamma^* u_{a, +} = 0.
\ee

The QCD interaction in terms of these component fields can be
reconstructed by gauge symmetry starting from the kinetic terms

\be \label{Skin}
S_{kin} =
-\int d^4x 2 A^{*a} (\D_+ \D_- - \D^{\perp} \D^{\perp *} ) A^{a}
\ee

The gauge group structure will be  written by using
brackets combining two fields into the colour states of the adjoint ($a$)
and  of the two fundamental ($\alpha$ and $*\alpha$) representations:
\bea (A_1^* T^a A_2 ) = -i f^{a b c } A_1^{* b} \ A_2^c, \
     (f_1^* t^a f_2) = t^a_{\alpha \beta} f^{* \alpha}_1 f_2^{\beta},
  \label{brackets}
\eea

We shall restrict the detailed discussion to pure gluodynamics.

The  interaction terms are recovered from the
kinetic term  written as

$$
   -2 A^{*a} \D_+ \D_-A^{a}   -
\D^{\perp} \D A^{* a} \D^{-2}  \D^{\perp *} \D A^{a}
$$
by extending the transverse derivatives in this  form,
   relying on the residual gauge
symmetry,
\be
\label{extend}
\D^{\perp} A^{a} \rightarrow ({\cal D}^{\perp } A)^{a} =
\D^{\perp } A^{a} + \frac{ig}{2} (A^* T^{a} A),
\ee
The result can be written as

$$ S= S_{kin} + S_3 + S_4, $$
\bea
S_3 \ = \  \frac{g}{2} \int d^4 x
   ( \partial_1^2 \hat V^*_{123}
  A^a (x_1) (A^*(x_2) T^a A^*(x_3)|_{x_i = x} + {\rm c.c.}),
\cr
S_4 \ = \  \frac{g^2}{4} \int d^4 x  
    \hat V_{1 1^{\prime}, 2 2^{\prime} }
 (A^*(x_1) T^c \partial A(x_{1^{\prime}})
(\partial A^*(x_2) T^c A(x_{2^{\prime}}) )
 |_{x_i = x_i^{\prime } = x } \ .
\label{action}
\eea
The elimination of redundant field components has lead to non-local
vertices,
\bea
\hat V^*_{123} = {i \over 3 \partial_1 \partial_2 \partial_3} [
\partial_{\perp 1}^* (\partial_2 - \partial_3) + {\rm cycl.} ], \ \
\hat V_{ 1 1^{\prime }, 2 2^{\prime }} = (\partial_1 +
\partial_{1^{\prime }})^{-2}.
\eea
Here and in the following we omit the space index $+$ on derivatives, i.e.
derivative operators not carrying subscripts $-, \perp $ are to be read as
$\partial_+$.  Integer number subscripts refer to the space point
on which the derivative  acts. The definition of the inverse
$\partial^{-1}$ is to be specified.

\begin{figure}[htb]
\begin{center}
\epsfig{file=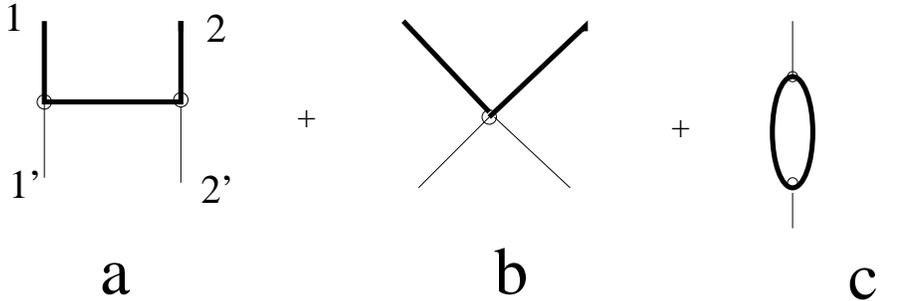,width=12cm}
\end{center}
\vspace*{0.5cm}
\caption{\small
Graphs contributing to the one-loop effective parton interaction. }
\end{figure}

Formally the effective parton interaction can be obtained by substituting
the field components as
\be
A = A_p + A_q + A_{p\p}
\ee
and integrating over the quantum fluctuations $A_q$ in logarithmic
approximation. In the resulting vertices involving the $A_p$ and
$A_{p\p}$ modes the intgration over the coordinates $x^{\perp}, x_+$ can
be done approximately since the dependence of these field modes on them
is specified.
We are interested in the vertices involving two  $A_p$ and two
$A_{p\p}$. They result from graphs Fig.1. We encounter two types of
integrals and extract their logarithmic contributions. The one with 3
propagators, Fig. 1a, results in
\bea \label{3prop}
\int \D_1^{\perp} \D_2^{\perp *} \ [x_{1 1\p}^{-2} x_{22\p}^{-2}
x_{1\p2\p}^{-2} ]
\rightarrow \ \ J_{111} \int {d^2 x^{\perp} \over |x^{\perp }|^2 } \cr
J_{111} = \int_0^1 d\alpha_1 d\alpha_2 d\alpha_3 \delta (\sum \alpha_i
-1) \delta(z_{11\p} - \alpha_1 z_{12})
 \delta(z_{22\p} + \alpha_2 z_{12})
\eea
The other intergral involves two propagators, Fig. 1b,
\bea \label{2prop}
\int  x_{1 1\p}^{-2} x_{22\p}^{-2} \delta(x_{1\p 2\p}^{\perp} )
\rightarrow \ \ J_{0} \int {d^2 x^{\perp} \over |x^{\perp }|^2 } \cr
J_{0} = \int_0^1 d\alpha_1 d\alpha_2  \delta (\sum \alpha_i
-1) \delta(z_{11\p} - \alpha z_{12})
 \delta(z_{22\p} + \alpha z_{12})
\eea

The result of the disconnected contributions, Fig. 1c,  has the form
\bea
\delta(z_{11\p}) \delta (z_{22\p}) C_p \left (\int_0^1 {d\alpha \over
\alpha } + w_p \right ) \cdot \int {d^2x^{\perp} \over |x^{\perp}|^2 }\cr
C_A=N, \ \ \ C_f ={N^2 -1 \over 2N}, \ \ \
w_A = - \frac{1}{12} ( 11- 2 {N_f \over N}), \ \ w_f = - \frac{3}{4}.
\eea

As an intermediate step we write the contributions  of Fig. 1 a, b,
where some terms are still not well defined because of infrared
divergencies.

 \bea
\int dz_1 dz_2 dz_{1^{\prime}} dz_{2^{\prime}}
\left [ {\partial_1^2 \partial_{2^{\prime}}^2 + \partial_{1^{\prime }}^2
\partial_2^2 \over (\partial_1 + \partial_{1^{\prime}} )^2  }
 J_{111} +
 {\partial_1 \partial_{2^{\prime}} + \partial_{1^{\prime }}
\partial_2 \over (\partial_1 + \partial_{1^{\prime}} )^2 }  \D_1 \D_2
 J_0  \right ]
\cr  (A_1^* T^{a} A_{1^{\prime}} )\
  (A_2^* T^{a} A_{2^{\prime}} ) \cr
+
\left [ {\partial_{1^{\prime }}^2 \partial_{2^{\prime }}^2 +
\partial_1^2 \partial_2^2 \over (\partial_1 + \partial_{1^{\prime}})^2
}
\  J_{111} +
 { (\partial_{1^{\prime }} \partial_{2^{\prime }} +
\partial_1 \partial_2)  \over
(\partial_1 + \partial_{1^{\prime}})^2 }
\D_1 \D_2 \  J_0 \right ]
(A_1^* T^{a} A_{1^{\prime }} ) \ (A_2 T^{a} A_{2^{\prime }}^* )  \cr
+ {(\partial_1 + \partial_{1^{\prime}} )^2 \over \D_1 \D_2} \  J_{111} \ \
(A_1^* T^{a} A_{1^{\prime }}^* ) \ (A_2 T^{a} A_{2^{\prime }} ) \cr
 -  { ( \partial_{1} \partial_{1^{\prime }} +
\partial_2 \partial_{2^{\prime }})  \over
(\partial_1  + \partial_{2})^2 }
\ \D_1 \D_2   J_0 \ \
(A_1^* T^{a} A_{2} ) \ (A_{1^{\prime }} T^{a} A_{2^{\prime }}^* )  \cr
\label{vertexa1}
\eea
We abbreviate the dependence of the parton fields on the light ray position by
subscripts, e.g. $A_1 = A (z_1) $. 
We sum the terms in the square brackets in order to cancel infrared
divergencies by applying the relation of $J$ integrals,
\bea
\label{Jf}
\D_1 \D_{2\p} J_{111} + \D_1 \D_2 J_0 = (\D_1 + \D_{1\p} )^2 \
\D_{1\p} \D_{2\p} \ J_{11\p}^{(1)} , \cr
J_{11\p}^{(p)} = - \int_0^1 d\alpha {(1-\alpha)^p \over \alpha }
\delta (z_{11\p} - \alpha z_{12}) \delta (z_{22\p} )
\eea
The first square bracket transforms to $\D_{1\p} \D_{2\p}
(J_{11\p}^{(2)} + J_{22\p}^{(2)} )$, where $J_{22\p} $ is obtained from
$J_{11\p} $ by the substitution $ 1 1\p \leftrightarrow 22\p $. The
disconnected contribution, Fig. 1c, amounts in adding
\be
2 \left  ( \int_0^1 {d\alpha \over \alpha } \ + \ w_g \right ) \delta
(z_{11\p} ) \delta (z_{22\p} )
\ee
to both square brackets resulting in improving $J_{11\p} + J_{22\p} $by
the standard + prescription. We assume it to be included in the
definition of $J_{11\p}$ in the following.

This transformation brings the first bracket in the final form. The
difference of the second from the first bracket,
$$ - (\D_{1\p} - \D_{1}) (\D_{2\p} -\D_{2}) J_{111} + \D_1 \D_2 J_0, $$
is then transformed by using the relations
\bea
\label{J112}
- ( \D_1 \D_{2\p} + \D_2 \D_{1\p} ) J_{111} = \D_{1\p} \D_{2\p} J_{112}
+ \D_1 \D_2 J_0,  \cr
(\D_1 + \D_{1\p} )^2 J_{111} = - \D_{1\p} \D_{2\p} J_{221}
- ( \D_1 \D_{2\p} + \D_2 \D_{1\p} )
{\D_1 \D_2 \over (\D_1 + \D_2)^2} J_{0}.
\eea
We use the notations
\bea
\label{JnJ0}
J_{n_1 n_2 n_3} =
{\Gamma(n_1+n_2+n_3-1) \over \Gamma(n_1) \Gamma (n_2) \Gamma (n_3) }
\int_0^1 d\alpha_1 d\alpha_2 d\alpha_3 \delta(\sum \alpha_i -1) \cr
\alpha_1^{n_1 -1} \alpha_2^{n_2 -1} \alpha_3^{n_3 -1} \ 
\delta ( z_{11\p} - \alpha_1 z_{12} )
\delta( z_{2 2\p}+ \alpha_2 z_{1 2} ),
\cr
J_0^{(p)} =
({\D_1 \D_2 \over (\D_1 + \D_2)^2})^p J_{0} = \cr
\int_0^1 d\alpha (\alpha (1-\alpha) )^p \delta (z_{11\p} - \alpha z_{12})
\delta (z_{22\p} + (1-\alpha) z_{12}).
\eea
The second relation in (\ref{J112})) is also applied to the third term
in (\ref{vertexa1}). Then one observes the cancellation of the last
term involving $J_0$ against other $J_0$ terms emerging from the
transformations by (\ref{J112}). The cancellation is due to the crossing
relation
\be \label{crossing}
(A_1 T^a A_{1\p}) (A_2 T^a A_{2\p} )  -
(A_1 T^a A_{2\p}) (A_2 T^a A_{1\p} )  =
(A_1 T^a A_{2}) (A_{1\p} T^a A_{2\p} )
\ee
implied by the Jacobi identity.

 The final form of the effective
gluon-gluon interaction vertices in the Bjorken asymptotics reads

\bea
 4 [  J_{1 1^{\prime }}^{(2)} + J_{22\p}^{(2)} + 2 w_g
\delta(z_{11\p} ) \delta (z_{22\p}) ] \cr
\{ (\D^{-1} A_1^*T^a \D A_{1\p}) (\D^{-1} A_2^* T^a \D A_{2\p} ) +
\ (\D^{-1} A_1^*T^a \D A_{1\p}) (\D^{-1} A_2 T^a \D A_{2\p}^* ) \} \cr
- 4 (J_{221} + 2 J_{112} )
 ( \partial^{-1} A_1^{*} T^{a} \partial A_{1^{\prime}})
( \partial^{-1} A_2 T^{a} \partial A^*_{2^{\prime }})   \cr
- 4   J_{221}
 ( \partial^{-1} A_1^{*} T^{a} \partial A_{1^{\prime}}^*)
( \partial^{-1} A_2 T^{a} \partial A_{2^{\prime }})
% +{[ A \leftrightarrow A^*]}
\label{hgg}
\eea

For later comparison we write also the result for the effective
fermion-fermion interaction (one flavour and one chirality only)

\bea
  [ J_{1 1^{\prime }}^{(1)} +J_{22\p}^{(1)} + 2 w_f
\delta (z_{11\p}) \delta (z_{22\p})  ] \cr
 \{ (\partial^{-1} f_1^* t^{a} f_{1^{\prime}} )
 ( \partial^{-1} f_2^* t^{a} f_{2^{\prime }})  +
 (\partial^{-1} f_1^* t^{a} f_{1^{\prime}} )
 ( \partial^{-1} f_2 t^{a} f_{2^{\prime }}^*) \}  \cr
- J_{111} \ \
 (\partial^{-1} f_1^* t^{a} f_{1^{\prime}} )
 ( \partial^{-1} f_2 t^{a} f_{2^{\prime }}^*)   \cr
- 2  J_0^{(1)}  \ \ \
(\partial^{-1} f_1^* t^{a} f_{2})
 (\partial^{-1} f_{1^{ \prime}} t^{a} f_{2^{\prime }}^*)
% + {[ f \leftrightarrow f^*]}
\label{hff}
\eea

\subsection{Conformal symmetry}

The conformal transformations of the light-ray position $z$
and of the parton fields $A_p(z), A_{p\p}(z)$ are based on the
Lie algebra $sl(2)$,
\be
  [S^0, S^{\pm} ] =  \pm S^{\pm}, \ \ \ [S^+, S^-] = 2 S^0
\ee
The action on the light ray, infinitesimal M\"obius transformations, are
generated by
\be
 S^{(0),-} = - \partial,  \ \ \ \  S^{(0) 0} = z \partial, \ \ \ \
S^{(0) +} = + z^2 \partial.
\ee
We need generic representations of conformal weight $\ell$
on functions of $z$ generated by
\be
 S^{(\ell ) -} = S^{(0) -}, \ \ \
S^{(\ell) 0 } = z^{-\ell} S^{(0) 0} z^{\ell},
\ \ \ S^{(\ell) +} = z^{-2 \ell } S^{(0) +} z^{2\ell}.
\ee
The representation space $V_{\ell}$ of single-parton wave functions
$\psi (z)$  is spanned by monomials $\psi^{(m)} = z^m$,
 the constant function
$\psi^{(0)}= 1$ represents the lowest weight state.
(In this convention the negative integer and half-integer values of $\ell$
correspond to the finite-dimensional representations related to angular
momentum and spin.)

Following \cite{Derkachov:1996ph},
in $V_{\ell}$ a scalar product is introduced, defined on the basis by
\be
<z^m, z^n >_{(\ell)} =
 \ \ \delta_{n,m} \ {\Gamma(m+1) \Gamma (2\ell)
\over \Gamma (m + 2\ell) }
 \ee
It obeys the symmetry condition
\be
< S^{0} \psi_1, \psi_2 >_{(\ell)} = <\psi_1, S^{0} \psi_2 >_{(\ell)}, \ \
<S^{+} \psi_1, \psi_2 >_{(\ell)} =
 - < \psi_1, S^{-} \psi_2 >_{(\ell)},
\ee

%provided
%\be
%a_{n,m}^{(\ell)} \ \ = \ \ \delta_{n,m} \ {\Gamma(m+1] \Gamma (2\ell)
%\over \Gamma (m + 2\ell) }
%\ee
The scalar product $< \psi_1, \psi_2 >_{(\ell)} $ can be calculated by the
action of the differential operator $\hat \psi_1 (\D ) $ on $ \psi_2 (z) $,
\be
 < \psi_1, \psi_2 >_{(\ell)} =
\hat \psi_1(\D) \ \psi_2 (z) |_{z=0},
\ee
where the symbol of this operator $\hat \psi (\zeta) $ is calculated
from the function $\psi(z) $ as
\be
\psi (z) \sum_m c_m z^m, \ \ \hat \psi(\zeta) = \sum_m c_m {\Gamma(2\ell)
\over \Gamma(m+2\ell)} \zeta^m .
\ee
%{\tt eventuell weglassen }
%The action os $S_{ell}^a $ on $\psi(z)$ is isomorphic to the action of
%$\hat S_{\ell}^a$ on the symbol $\hat \psi (\zeta)$, where
%\be
%\label{Saisom}
%\hat S_{\ell}^+ = \zeta, \ \
%\hat S_{\ell}^0 = \zeta \D_{\zeta} + \ell, \ \
%\hat S_{\ell}^- = - (\zeta \ \D_{\zeta}^2 + 2 \ell \D_{\zeta} ).
%\ee

The conformal symmetry properties of the multi-parton states
$\psi(z_1, ...z_k)$ are represented
on the tensor product $\otimes_1^k V_{\ell_i} $ and  the generators (at
tree level ) are
\be
S_{\{\ell_i\}}^a = \sum_{i = 1}^k S_{\ell_i}^a.
\ee
Irreducible representations of weight $L = \sum_1^k \ell_i + n, n= 0,1,
... $ are spanned by
\be
\psi^{(m)}_{ \{\ell_i\},n} =  ( S_{ \{\ell_i\} }^+ )^m \
\psi^{(0)}_{ \{\ell_i\},n}
\ee
with the lowest weight states obeying
\be
S_{ \{\ell_i\}}^0 \ \psi^{(0)}_{ \{\ell_i\},n} = (\sum_1^k \ell_i + n )
\psi^{(0)}_{ \{\ell_i\},n}; \ \ \ \
 S_{ \{ \ell_i\}}^- \  \psi^{(0)}_{ \{\ell_i\},n} \ = \ 0.
\ee
They are represented by a homogeneous polynomials of degree $n$ of the
differences $z_{ij} = z_i - z_j $.
The symmetry isomorphism between the wave functions $\psi(z_1, ...,z_k)$ and
the differential operators $\hat \psi (\D_1,..., \D_k)$ induced by the
scalar
product allows to formulate the symmetric relation between the
multiparton
states and the composite operators  of the parton
fields. If the field operators $A_i$ transform as conformal primaries of weight
$\ell_i$ then the composite field operators
\be
\label{compop}
\hat {\cal O} (z) = \hat \psi (\D_1, ...,\D_k) \ \ \prod_1^k A_i (z_i) \
|_{z_i \rightarrow z}
\ee
mix under one-loop renormalization only for those
 $\psi $ belonging to one particular
conformal symmetry representation  weight $L$ and one particular level 
$m$.
 With the lowest weight state,
i.e. specifying to $\hat \psi^{(0)}_{\{\ell_i\},n} $, the corresponding
operator on tree level transforms as the conformal primary of weight
$L = \sum \ell_i + n $. The relevant QCD bare conformal primaries
are the fermion field components $f, f^*, \tilde f, \tilde f^*$
with weight $\ell= 1$ and
the gluon field strength components $\D A, \D A^* $ with weight $\ell =
\frac{3}{2} $.
In the case of two partons, $ k= 2$, there is just one irreducible
representation for each $n$ and therefore
$\psi^{(m)}_{\ell_1,\ell_2,n} $ are eigenstates of the evolution at one loop
and the corresponding operators are multiplicatively renormalized.

Operators acting on two-parton states as $ \hat K \psi (z_1,z_2) =
\tilde \psi (z_1,z_2), \ \ \
\hat K: V_{\ell_{1\p}} \otimes V_{\ell_{2\p}} \rightarrow V_{\ell_1} \otimes
V_{\ell_2}
$,
are conformal symmetric if
\be
 (S_{\ell_1}^a + S_{\ell_2}^a ) \hat K \ = \ \hat K (S^a_{\ell_1\p} +
S^a_{\ell_2\p} )
\ee
Representing $\hat K$ in integral form
\be
\label{intform}
\hat K \psi (z_1, z_2) = \int_{C_{1\p},C_{2\p}}d z_{1\p} d z_{2\p}
K(z_1,z_2; z_{1\p}, z_{2\p} ) \ \psi ( z_{1\p}, z_{2\p} )
\ee
the symmetry condition on the kernel reads
\be
\label{symcond}
\left ( S_{\ell_1,1}^a + S_{\ell_2,2}^a -(S_{\ell_1\p,1\p}^a)^T
- (S_{\ell_2\p,2\p}^a)^T \right )  \
K(z_1,z_2; z_{1\p}, z_{2\p} ) \ = \ 0.
\ee
In the case of integration over closed contours the transposed generators
are simply
\be
\label{lbar}
(S_{\ell}^a )^T = - S_{\bar \ell}^a , \ \ \ \bar \ell = 1 - \ell,
\ee
and the kernel is a conformal symmetric 4-point function,
\be
K_{(\ell_1,\ell_2,\bar \ell_{1\p}, \bar \ell_{2\p})}
(z_1,z_2;z_{1\p}, z_{2\p} )
\sim
< \  A_{\ell_1} (z_1) \ A_{ \ell_2} (z_2)  \ A_{\bar \ell_{1\p}}
(z_{1\p}) \ A_{\bar \ell_{2\p}} (z_{2\p}) >,
\ee
where $ A_{\ell} $ stands for a conformal primary field operator of
weight $\ell$. A generic solution of the symmetry condition 
(\ref{symcond}) reads
\be
\label{kerneldh}
K_{(\ell_1,\ell_2,\bar \ell_{1\p},\bar \ell_{2\p})}^{(d,h)} (z_1,z_2;z_{1\p},
z_{2\p} )
= z_{12}^{a_{12}} \ z_{1\p 2\p}^{a_{1\p 2\p}}
z_{11\p}^{a_{11\p}} \ z_{2 2\p}^{a_{2 2\p}}
z_{12\p}^{a_{12\p}} \ z_{2 1\p}^{a_{2 1\p}}
\ee
with the exponents depending on the weights and on two parameters $d, h$
as
\bea
\label{expa}
a_{12} = d + \frac{1}{2} \sigma - \ell_1 -\ell_2,
a_{1\p2\p} = d + \frac{1}{2} \sigma - \bar \ell_{1\p} -\bar \ell_{2\p},
\cr
a_{12\p} = h + \frac{1}{2} \sigma - \ell_1 - \bar \ell_{2\p},
a_{2 1\p} = h + \frac{1}{2} \sigma  -  \ell_{2} -\bar \ell_{1\p}, \cr
a_{11\p} = -d -h - \ell_1 - \bar \ell_{1\p},
a_{2 2\p} = -d -h -  \ell_{2} -\bar \ell_{2\p}, \cr
\sigma = \ell_1 + \ell_2 + \bar \ell_{1\p} + \bar \ell_{2\p}.
\eea
Any symmetric 4-point function is a sum of such expressions with
coefficients depending on the parameters $d, h$; it differs from
(\ref{kerneldh}) by a factor being a function of the anharmonic ratio
$r_{st} = {z_{11\p} z_{22\p} \over z_{12} z_{1\p 2\p} }$. In (\ref{kerneldh})
the powers with exponents $d, h$ can be written as the powers of the
anharmonic ratios
$
(r_{st})^{-d}  (r_{ut} \ r_{st}^{-1} )^{h}
$
which are not independent,
\be
\label{anharm}
r_{ut} = {z_{12\p} z_{2 1\p} \over z_{12} z_{1\p 2\p} }
\ = \  r_{st} - 1.
\ee

In view of the branch points appearing for non-integer $a_{ij}$ we specify
closed integration contours in (\ref{intform}) as the Pochhammer
contours
$C_{2\p} = [2^+, 1^{\prime +} ,2^-,1^{\prime -}]$, $ C_{1\p} = [ 2^+,
1^+, 2^-, 1^-] $, where e.g. $2^+$ means to go around the point $z_2  $
in positive sense.

Actually we should have the integration running over the light ray, i.e. the
real axis. In the regular case, $a_{ij} > -1$, the contours can be contracted
easily to the real axis on which $z_1, z_2$ are located. It is convenient to
write the resulting kernel as
\bea
\label{alphakernel}
z_{12}^{2-\sigma} \int_0^1 d\alpha_1 d\alpha_2 d\alpha_3
\delta (\sum \alpha_1 -1)
\alpha_1^{a_{11\p} } \alpha_2^{a_{22\p}} \alpha_3^{a_{1\p 2\p}}
(1-\alpha_1)^{a_{21\p}}  (1- \alpha_2)^{a_{22\p}} \cr
\delta (z_{11\p} - \alpha_1 z_{12} )  \delta (z_{22\p} + \alpha_2 z_{12} )
\eea
The contraction has to be done with more care in  the singular cases.
For example, if one exponent goes to $-1$ the result of the contraction is
reproduced by the substitution in (\ref{alphakernel})
\bea
\label{singul}
\alpha^{-1 + \varepsilon} \rightarrow  {1 \over \varepsilon}\delta(\alpha) +
 {1 \over [\alpha]_+ }, \cr
\int_0^1 d\alpha  {1 \over [\alpha]_+ } \phi(\alpha) \ = \
\int_0^1 d\alpha  {1 \over \alpha } (\phi(\alpha) - \phi(0)).
\eea

Consider the case of equal weights, $\ell_1 = \ell_2 = \ell_{1\p} = \ell_{2\p} =
\ell $. An example of singular contraction contributions arises for
the parameters $d= \varepsilon \rightarrow 0, h= 0$. We obtain the kernel of
the identity operator as the leading contribution in $\varepsilon$. The
next term is proportional to
$
J_{11\p}^{(2\ell-1)} + J_{22\p}^{(2\ell-1)}
$, with the notation defined in (\ref{Jf}). For the parameters
$h=0$ and $  -2\ell < d < 0 $ (assuming $\ell > 0$) we have the regular
case of
contraction and obtain the kernel $J_{-d,-d, d+2\ell } $, with the notation
given in (\ref{JnJ0})). Another singular case of contraction arises at
$d = - 2 \ell +
\varepsilon $, the leading term resulting in $J_0^{(2\ell -1)}$, with the
notation given in (\ref{JnJ0})).
The latter examples cover the kernels encountered in the QCD result
above, for the particular parameter values
 $$ h= 0,\ \ \  d = -\varepsilon,  \ \ d = -1,  \ \ d= -2 (+ \varepsilon) $$
and with the conformal weights $\ell = \frac{3}{2}$ for the gluon-gluon
vertices and $\ell = 1$ for the quark-quark vertices.

Also the kernels appearing in the mixed gluon-quark and annihilation-type
vertices can be identified, with the appropriate choice of the weights,
 as particular cases of conformal 4-point functions \cite{DKBj}.

Generating  the evolution in the Bjorken limit the effective vertices
(\ref{hgg}) can be considered as second-quantized parton field operators by
reading $A_{p\p}$ as annihilation and $A_p$ as creation operators with the
contraction rules $$ < A^*_1 (z_1) \ A_{1\p} (z_{1\p}) > =
< A_1 (z_1) \ A^*_{1\p} (z_{1\p}) > = \frac{1}{2} \delta (z_{11\p}). $$
Two parton states onto which these operators act
are defined by  the wave function, e.g. for two gluons of helicities
$+ 1 $ and $-1$
\be
\label{pstates}
| n, m; A^*, A > \ = \ \int_{-\infty}^{+\infty} dz_1 dz_2
\psi_n^{(m)} (z_1, z_2) \
\D^{-1} A_1^* (z_1)  \D^{-1} A_2 (z_2) \ |0>
\ee
The label $A$ stands for parton type (gluon) and helicity (-1).
For representing the action of the vertices on the composite operators
(\ref{compop}) we read the involved parton fields as annihilation operators
$(A_{p\p})$. It is clear that the eigenstates have the form
$| n, m; A^*, A > \pm | n. m; A, A^* >,
| n, m; A^*, A* >,   | n. m; A, A > $. They contribute, respectively,
 to the $t$ channel exchange in cases of
unpolarized, polarized (helicity asymmetry) and
transversity (generalized) parton distributions.

%%%%%%%%%%%%%%%%%%%%%%%%%%%%%%%%%%%%%%%%%%%%%%%%%%%%%%%%%%%%%%%%%%%%

\section{Parton interaction in SYM}
\setcounter{equation}{0}

The kinetic terms of the super Yang-Mills theory are obtained from the
ones of QCD  (\ref{Skin}) by changing the fermion representation to the
adjoint ($f^{\alpha} \rightarrow f^a, t^{\alpha} \rightarrow T^a $) and
by restricting to one
flavour and one chiral component (i.e. omitiing the $\tilde f $ term).
We introduce superfields,
\be
\label{superfields}
{\cal A }^a (x, \theta) = A^a (x) + c\  \theta f^a (x) , \ \
{\cal A }^{* a} = A^{ * a} + c \  \theta f^{* a}, \ \
c = {1 \over \sqrt {2i}}
\ee
in order to write the kinetic terms as
\bea
\label{SkinSYM}
S^{SYM}_{kin} = -2 \int d^4 x d \theta \
{\cal A}^{*
a} \D^{-1} D (\D_+ \D_- - \D^{\perp} \D^{* \perp} ) \ {\cal A}^a \cr
D = \D_{\theta} + \theta \D, \ \ D^2 = \D.
\eea
This action is invariant with respect to super translations generated
by $S^{-\frac{1}{2}} = - \D_{\theta} + \theta \D $.
The interaction terms can be reconstructed by super gauge symmetry, i.e.
by extending the transverse derivatives by
\be
\label{superextend}
\D^{\perp} {\cal A}^{a} \rightarrow ({\cal D}^{\perp } {\cal A})^{a} =
\D^{\perp } {\cal A}^{a} + \frac{ig}{2} ({\cal A}^* T^{a} A),
\ee
and replacing the integrand in the kinetic term as
\be
 {\cal A}^{*
a}  D  \D_-   \ {\cal A}^a -
({\cal D}^{*\perp} D {\cal A}^* )^a
 \D^{-2} D  ({\cal D}^{\perp} D  {\cal A})^a.
\ee
The result is
$$ S^{SYM}_{kin} + S_3^{SYM} + S_4^{SYM}, $$
\bea
\label{superaction}
S_3^{SYM} \ = \  \frac{g}{2} \int d^4 x d \theta
   ( \partial_1 \hat V^*_{123}
 D_1 {\cal A}^a_1  ({\cal A}^*_2 T^a {\cal A}^*_3 |_{(x_i,\theta_i) =
(x,\theta )} +
{\rm c.c.}), \cr
S_4^{SYM} \ = \  \frac{g^2}{4} \int d^4 x d \theta
     ({\cal A}^*_1 T^c D {\cal A}_{1^{\prime}} )  \ \D^{-2} D \
(D {\cal A}^*_2 T^c {\cal A}_{2^{\prime}} )
 |_{(x_i,\theta_i) = (x, \theta)} \ .
\label{action}
\eea
We have abbreviated the dependence of the superfields on spacetime position
$x_i$ and $\theta_i$ by subscript $i = 1,2,1\p, 2\p$.

The close similarity to the gluodynamic case allows to follow the
previous calculation with minor modifications. The calculation in terms
of superfields preserves the symmetry with respect to the super
translation $S^{-\frac{1}{2}} = - \D_{\theta} + z \D $ 
in all steps. We use the super propagators
\bea
< {\cal A}^*_1 \ {\cal A}_2 > \sim (x_{+ 12} \tilde z_{12} -
|x_{12}^{\perp}|^2 )^{-2}, \cr
< {\cal A}^*_1 \ \D^{-1} D {\cal A}_2 > \sim     \theta_{12}
(x_{+ 12}  z_{12} - |x_{12}^{\perp}|^2 )^{-2}.
\eea
$$ x_{12} = x_1 - x_2, \tilde z_{12} = z_1 - z_2 - \theta_1 \theta_2,
\theta_{12} = \theta_1 - \theta_2. $$

The resulting effective vertices can be represented in a form close to
the one of $S_4^{SYM}$. The result for Fig. 1 a,b is

 \bea
\int dz_1 dz_2 dz_{1^{\prime}} dz_{2^{\prime}} d\theta_1 d \theta_2
d\theta_{1\p} d \theta_{2\p} \cr
\left [ {\partial_1 D_1 \partial_{2^{\prime}} D_{2\p} +
\partial_{1^{\prime}}  D_{1\p}  \partial_2 D_2
\over (\partial_1 + \partial_{1^{\prime}} )^2 }
 \theta_{11\p } \theta_{22\p} \tilde J_{111} +
 {D_1 D_{2^{\prime}} + D_{1^{\prime }}
D_2 \over (\partial_1 + \partial_{1^{\prime}} )^2 }    \D_1 \D_2
 \theta_{11\p} \theta_{22\p} \tilde J_0  \right ]
\cr  ({\cal A}_1^* T^{a} {\cal A}_{1^{\prime}} )\
  ({\cal A}_2^* T^{a} {\cal A}_{2^{\prime}} ) \cr
+
\left [ {\partial_{1^{\prime }} D_{1\p} \partial_{2^{\prime }} D_{2\p} +
\partial_1 D_1 \partial_2 D_2 \over (\partial_1 +
\partial_{1^{\prime}})^2 }
\  \theta_{11\p} \theta_{22\p} \tilde J_{111} +
 { D_{1^{\prime }} D_{2^{\prime }} +
D_1 D_2  \over
(\partial_1 + \partial_{1^{\prime}})^2 }
\D_1 \D_2 \  \theta_{11\p} \theta_{22\p} \tilde J_0 \right ] \cr
({\cal A}_1^* T^{a} {\cal A}_{1^{\prime }} ) \ ({\cal A}_2 T^{a}
{\cal A}_{2^{\prime }}^* ) \cr
 - {(D_1 + D_{1^{\prime}} ) ( D_2 + D_{2\p} ) } \  \theta_{11\p}
\theta_{22\p} \tilde J_{111}
\ \ ({\cal A}_1^* T^{a} {\cal A}_{1^{\prime }}^* ) \ ({\cal A}_2 T^{a}
{\cal A}_{2^{\prime }} ) \cr
 -  { ( D_{1} D_{1^{\prime }} +
D_2 D_{2^{\prime }})  \over
(\partial_1  + \partial_{2})^2 }
\ \D_1 \D_2   \theta_{12} \theta_{1\p 2\p} \tilde J_0 \ \
({\cal A}_1^* T^{a} {\cal A}_{2} ) \ ({\cal A}_{1^{\prime }} T^{a}
{\cal A}_{2^{\prime }}^* )
\label{vertexsym}
\eea
Here the ${\cal A}_i$ denote the parton super field with the subscript
abbreviating the argument $(z_i, \theta_i)$. 
 The kernels with tilde $\tilde J$ are obtained from
the corresponding kernels $J$ defined in sect. 2  
by replacing the distances $z_{11\p} ,
z_{22\p}, z_{1\p 2\p}$ by their supersymmetric extensions $\tilde
z_{11\p} = z_{11\p} - \theta_1 \theta_{1\p}, \tilde z_{22\p},
\tilde z_{1\p 2\p} $.
The integrand of the SYM results differs from the one of the gluodynamic
result by merely the latter replacement, factors $\theta_{11\p}
\theta_{22\p} $ or $\theta_{12} \theta_{1\p 2\p}$ and the replacement of
some derivatives $\D$ by $D$.

\section{Superconformal symmetry}
\setcounter{equation}{0}

The superconformal transformations of the extended light ray ($z,
\theta $) and of the superfields residing on it relies on the algebra
$osp(2|1), S^a, a= 0, \pm \frac{1}{2}, \pm 1$,
\bea
\label{osp}
[S^0, S^a] = a S^a, \ \ \  [S^{+1}, S^{-1}] = 2 S^0, \ 
[S^{-\frac{1}{2}}, S^{+1}] =  S^{+\frac{1}{2}},              \
[S^{+\frac{1}{2}}, S^{-1}] =  S^{-\frac{1}{2}}, \cr
[S^{-\frac{1}{2}}, S^{-\frac{1}{2}}]_+ = 2 S^{-1}, \ 
[S^{+\frac{1}{2}}, S^{+\frac{1}{2}}]_+ = - 2 S^{+1}, \ 
[S^{+\frac{1}{2}}, S^{-\frac{1}{2}}]_+ = - 2 S^{0}
\eea
From the last 3 relations we see that the algebra is generated by
$S^{\pm \frac{1}{2}}$. Super M\"obius transformations of the extended
light ray are generated by the particular representation $S_0^{a}, \ell
= 0$, where
\be
S_0^{-\frac{1}{2}} = - \D_{\theta} + \theta \D, \ \ \
S_0^{+\frac{1}{2}} = z \ S_0^{-\frac{1}{2}},
\ee
and the remaining generators following from the last 3 commutation
relations (\ref{osp}). Representations on the fields are labeled by the
weight $\ell$ and are generated by $S_{\ell}^a$,
\be
\label{Sell}
S_{\ell}^{-\frac{1}{2}} = S_0^{-\frac{1}{2}}, \ \ \ \
S_{\ell}^{+\frac{1}{2}} = z^{-2\ell} \ S_0^{+\frac{1}{2}} \ z^{2 \ell}.
\ee
The corresponding representation space of 1-parton states, $\psi(z,
\theta) \in V_{\ell} $, is spanned by $z^m, \theta z^m, m= 0, 1,... $.
The constant function represents the lowest weight state.
A scalar product on $V_{\ell}$ with the symmetry properties
\bea
< S^{0} \psi_1, \psi_2 >_{(\ell)} = <\psi_1, S^{0} \psi_2 >_{(\ell)}, \ \
\cr
<S^{a} \psi_1, \psi_2 >_{(\ell)} =
 - < \psi_1, S^{a} \psi_2 >_{(\ell)},\ \ \  a= \pm \frac{1}{2}, \pm 1.
\eea
is defined on the basis by
\bea
\label{product}
<z^m, z^n >_{(\ell)} =
 \ \ = \ \ \delta_{n,m} \ {\Gamma(m+1) \Gamma (2\ell)
\over \Gamma (m + 2\ell) }
\cr
<\theta z^m, \theta z^n >_{(\ell)} =
 \ \ = \ \ \delta_{n,m} \ {\Gamma(m+1) \Gamma (2\ell)
\over \Gamma (m+1 + 2\ell) } \cr
<\theta z^m,  z^n >_{(\ell)} =
< z^m, \theta z^n >_{(\ell)} = 0.
\eea
It can be calculated by differentiations,
\be
< \psi_1, \psi_2 >_{(\ell)} =  \hat \psi_1 (\D, D) \ \psi_2 (z,
\theta)|_{z=0, \theta = 0},
\ee
where the symbol of the differential operator $\hat \psi (y, \vartheta
)$ is calculated from the expansion of the function $\psi(z, \theta)$,
\bea
\psi (z, \theta) = \sum_m (a_m + \theta b_m) z^m, \cr
\hat \psi (y, \vartheta ) = \Gamma (2 \ell )
\sum_m  ( a_m + \vartheta  { b_m \over 2 \ell + m }  )
{ 1 \over \Gamma (m + 2\ell) } y^m
\eea
The action of $S_{\ell}^a$ on $\psi (z, \theta )$ is isomorphic to the
action of $\hat S_{\ell}^a $ on $\hat \psi (y, \vartheta )$, where
\be
\hat S_{\ell }^{-\frac{1}{2}} = - y \D_y \D_{\vartheta} + \vartheta \D_y -
2\ell \D_{\vartheta}, \ \ \
\hat S_{\ell }^{+\frac{1}{2}} =   \vartheta -  y \D_{\vartheta},
\ee
and $\hat S_{\ell}^a$ obey (\ref{osp}).

Multi ($k$) parton states are represented by functions of $(z_i,
\theta_i), i= 1, .., k $, $\psi (z_1, \theta_1, ..., z_k, \theta_k) \in
\otimes_1^k V_{\ell_i} $
and the transformations are generated by the sum
$\sum_{i=1}^k S_{\ell_i}^a = S_{\{\ell_i \}}^a $.

The lowest weight states of the representations into which the tensor
product decomposes obey
\bea
 S_{\{\ell_i \} }^{-\frac{1}{2} }
 \psi^{(0)}_{ \{ \ell_i \},n} (z_1, \theta_1, ..., z_k, \theta_k)  = 0,
\ \ \  S_{\{ \ell_i\} }^{0}
 \psi^{(0)}_{\{\ell_i\},n} = \
(\sum \ell_i + n) \psi^{(0)}_{\{\ell_i\},n}, \cr
n \ = \  0, \frac{1}{2}, 1, ...
\eea
and, therefore, are homogeneous functions of $z_i, \theta_i$ of degree
$n$ of the super-translation invariant combinations
\be
\tilde z_{ij} = z_i - z_j - \theta_i \theta_j, \ \
\theta_{ij} = \theta_i - \theta_j
\ee
In the particular case of $k=2$ we have with the notations
$\bar n = $ integer part of $n$ and $ \nu = 2 (n - \bar n)$
\be
\label{lowest2}
\psi_{\ell_1,\ell_2,\bar n, 0}^{(0)} = \tilde z_{12}^{\bar n}, \ \ \
\psi_{\ell_1,\ell_2,\bar n, 1  }^{(0)} =
\theta_{12}  z_{12}^{\bar n}.
\ee
and the irreducible representations are spanned by
\be
\label{psim}
\psi_{\ell_1,\ell_2,\bar n, \nu}^{(m)} =
(S_{\ell_1}^{+\frac{1}{2}} + S_{\ell_2}^{+\frac{1}{2}} )^m \
        \psi_{\ell_1,\ell_2,\bar n, \nu}^{(0)}, \ \ \
m= 0, \frac{1}{2}, 1,...
\ee
in particular, for the next-to-lowest level $m = \frac{1}{2}$ we have for
$\ell_1 = \ell_2 = \ell $,
\be
\label{psi12}
\psi_{\ell, \ell,\bar n , 0}^{(\frac{1}{2}) } = (\bar n + 2 \ell)
(\theta_1 + \theta_2) z_{12}^{\bar n}, \ \ \
\psi_{\ell, \ell, \bar n ,1 }^{(\frac{1}{2})}
= - z_{12}^{\bar n +1} - (\bar n + 4 \ell) \theta_1 \theta_2
z_{12}^{\bar n}. \ee
The relation between parton states $\psi $ and composite operators mentioned
above generalizes to the supersymmetric case,
\be
\hat {\cal O} (z, \theta) = \hat \psi (\D_1,D_1, ...,\D_k,D_k)
\ \ \prod_1^k {\cal A}_i (z_i, \theta_i) |_{z_i, \theta_i) \rightarrow
(z, \theta) }
\ee

Also the notion of symmetric operators $\hat K$
carries over from the case of $sl(2)$ to the case of $osp(2|1)$.
If $\hat K$ is given in integral form
\bea
\label{intforms}
\hat K \psi (z_1,\theta_1,  z_2, \theta_2) = \cr
\int_{C_{1\p},C_{2\p}}d z_{1\p} d z_{2\p} d\theta_{1\p} d\theta_{2\p}
K(z_1, \theta_1, z_2, \theta_2 ; z_{1\p}, \theta_{1\p}, z_{2\p},
\theta_{2\p} ) \ \psi ( z_{1\p}, \theta_{1\p} z_{2\p}, \theta_{2\p} )
\eea
then the symmetry conditions on the kernel is analogous to
(\ref{symcond} ), where here the particular conditions with
$a= \pm \frac{1}{2} $ are sufficient.
For closed contours (in $z_{1\p}, z_{2\p}$ the transposition acts on the
generators as
\be
(S_{ \ell }^a )^T = - S_{\bar \ell}^a , \ \ \
\bar \ell = \frac{1}{2} - \ell,
\ee
where the relation between the weights $\ell, \bar \ell$ differs from the
non-supersymmetric case (\ref{lbar}). The particular condition on the
kernel with $a=
-\frac{1}{2}$ implies that the kernel $ K$ depends on the supertranslation
invariant combinations $\tilde z_{ij}, \theta_{ij}, (i,j = 1,2,1\p,2\p) $
only. The remaining independent condition with $a = + \frac{1}{2}$ is
solved by an expression 
$\tilde K_{\ell_1,\ell_2,\bar \ell_{1\p}, \bar \ell_{2\p}}^{(d, h)} $
of the form (\ref{kerneldh}) depending on  two
parameters $d,h,$ and on the weights $\ell_1, \ell_2, \ell_{1\p}, \ell_{2\p}
$ in the same way (\ref{expa}) but with $z_{ij} $
replaced by $\tilde z_{ij}$.
The powers with the exponents $d, h$ can be written as powers of the
anharmonic ratios likewise,
\be
\tilde r_{st} = {\tilde z_{11\p} \tilde z_{22\p} \over \tilde z_{12}
\tilde z_{1\p 2\p} }, \ \
\tilde r_{ut} = {\tilde z_{12\p} \tilde z_{2 1\p} \over \tilde z_{12}
\tilde z_{1\p 2\p} }
\ee
Two symmetric 4-point functions of the same set of weights differ by a
factor being a function of invariants built out of the coordinates of the 4
points in complete analogy to the $sl(2)$ case. The difference is that now
out of the coordinates of 4 points $(z_i, \theta_i)$ one can construct two
independent invariants. In particular the two anharmonic ratios
$\tilde r_{st}, \tilde r_{ut} $ are not dependent (unlike (\ref{anharm})),
but obey
\bea
\label{anharms}
\tilde r_{st} - \tilde r_{ut} - 1 = r_{\theta t}, \cr
r_{\theta t} = {1\over \tilde z_{12} \tilde z_{1\p 2\p}}
\left ( \tilde z_{22\p} \theta_{12\p} \theta_{11\p} +
\tilde z_{11\p} \theta_{21\p} \theta_{22\p} +
\tilde z_{1\p2\p} \theta_{11\p} \theta_{22\p} \right ).
\eea
We conclude that the from (\ref{kerneldh}) (with $z_{ij}$ replaced by
 $\tilde z_{ij}$) represents a generic form of ($osp(2|1)$)
superconformal 4-point functions. Sums with varying values of $d,h$ represent
all of them. Alternatively, one can consider
\be
\left (A_d  + B_{d}  \ \ r_{\theta t} \right )
\tilde K_{\ell, \ell_2, \bar \ell_{1\p}, \bar \ell_{2\p}}^{(d_1,0)}
\ee
as the generic form and represent arbitrary symmetric 4-point functions as
sums of them with varying parameter $d$.

For our applications we may
restrict ourselves to the case $ \ell_1, \ell_2, \bar \ell_{1\p},
\bar \ell_{2\p} = \ell = 1$. The basic two-parton wave functions $\psi_{\ell,
\ell, \bar n,\nu}^{(m)} $ are eigenfunctions
\bea
\int_{C_{1\p},C_{2\p}}d z_{1\p} d z_{2\p} d\theta_{1\p} d\theta_{2\p}
\left ( A_d + B_d r_{\theta t} \right ) \ \tilde  K_{(\ell)}^{(d,0)}  \ \ 
\psi_{\ell, \ell, \bar n,\nu}^{(m)} = \cr
(A_d \  \lambda_{\ell,\bar n, \nu}^{(d, 0)} + B_d
\lambda_{\ell,\bar n, \nu}^{(d,\theta)} ) \ \
\psi_{\ell, \ell, \bar n,\nu}^{(m)}
\eea
The eigenvalues do not depend on the level $m$.
\bea
\label{eigenv}
\lambda_{\ell,\bar n, \nu}^{(d, 0)} = (-1)^{\nu}
{\Gamma(d + \bar n + 2\ell +\frac{1}{2} ) \over
\Gamma(- d + \bar n + 2\ell +\frac{1}{2} ) } \Gamma (-d+\frac{1}{2} )^2,
\cr
\lambda_{\ell,\bar n, \nu}^{(d, \theta) }  =
{\Gamma(d + \bar n + 2\ell -\frac{1}{2} + \nu ) \over
\Gamma(- d + \bar n + 2\ell +\frac{1}{2} + \nu ) }
\Gamma (-d+\frac{1}{2} )^2.
\eea
The dependence on $\bar n, \nu$ can be calculated before the contours
have been contracted to the real axis by using the contour integral
representation of the Beta function.
In the regular case the contraction results in the kernels
\bea
\tilde K_{(\ell)}^{(d,0)} =
 \int_0^1 d\alpha_1 d\alpha_2 d\alpha_3 \delta(\sum \alpha_i -1)
\alpha_1^{-\frac{1}{2} - d }  \alpha_2^{-\frac{1}{2} - d}
\alpha_3^{d+ 2\ell - \frac{1}{2}},
 \cr
\delta(\tilde z_{11\p} - \alpha_1 [\tilde z_{11\p}+\tilde z_{1\p2\p} +
\tilde z_{2\p 2} ] )
\delta(\tilde z_{22\p} + \alpha_2 [\tilde z_{11\p}+\tilde z_{1\p2\p} +
\tilde z_{2\p 2} ] ) \cr
\tilde z_{12}^{d - 2 \ell + \frac{1}{2}}
[\tilde z_{12} + \theta_{1 2\p} \theta _{2 1\p} ]^{-d + 2 \ell + \frac{1}{2}}
\cr
r_{\theta t} \tilde K_{(\ell)}^{(d,0)} =
 \int_0^1 d\alpha_1 d\alpha_2 d\alpha_3 \delta(\sum \alpha_i -1)
\alpha_1^{-\frac{1}{2} - d}  \alpha_2^{-\frac{1}{2} - d}
\alpha_3^{d+ 2\ell - \frac{3}{2}}
 \cr
\delta(\tilde z_{11\p} - \alpha_1 [\tilde z_{11\p}+\tilde z_{1\p2\p} +
\tilde z_{2\p 2} ] )
\delta(\tilde z_{22\p} + \alpha_2 [\tilde z_{11\p}+\tilde z_{1\p2\p} +
\tilde z_{2\p 2} ] ) \cr
[\alpha_1 \theta_{2 1\p} \theta_{22\p} + \alpha_2 \theta_{11\p} \theta_{12\p}
+ \alpha_3 \theta_{11\p} \theta_{22\p} ]
\eea
For real $\ell > 0$ the regularity condition is $ \frac{1}{2} > d >
\frac{1}{2} - 2 \ell $ .
The kernel of the identity operator $ \delta (z_{11\p }) \theta_{11\p} \delta
(z_{22\p}) \theta_{22\p} $ is in $ r_{\theta t} \tilde K_{\ell}^{(d,0)}$ as the
leading term in $\varepsilon$ for $d \rightarrow +\frac{1}{2} -
\varepsilon $. The next term $\sim  \varepsilon^{-1}$ is proportional to
\bea
\label{JT}
J_T^{(\ell)} = J_{T1}^{(\ell)} -  J_{T2}^{(\ell)} \cr
J_{T1}^{(\ell)} =
- \int_0^1 d\alpha
\left [ {1-\alpha)^{2\ell} \over [\alpha]_+} \theta_{11\p} \theta_{22\p}
+ (1-\alpha)^{2\ell -1} \theta_{21\p} \theta_{22\p} \right ] \cr
\delta (\tilde z_{11\p} - \alpha [\tilde z_{11\p} + \tilde z_{1\p2\p} + \tilde
z_{2\p2} ]) \delta (\tilde z_{22\p}) 
\eea
$J_{T2}$ is obtained from $J_{T1}$ by the interchange of subscripts
$ 11\p \leftrightarrow 22\p  $.
For $d\rightarrow \frac{1}{2} - 2 \ell $ we find both singular and regular
contraction contributions in the decomposion with respect to the odd ($\theta$)
coordinates; e.g.
\be
\label{JV}
J_V^{(\ell)} = \lim_{\varepsilon \rightarrow 0} \ 2 \tilde 
K_{(\ell)}^{(\frac{1}{2} - 2 \ell + \varepsilon, 0)}
\ee
involves in the term proportional to $\theta_{1\p} \theta_{2\p}$
a contribution from singular contraction (by applying (\ref{singul}))
finite in the limit, whereas the terms in the remaining $\theta $ components
are regular contraction contributions.

%%%%%%%%%%%%%%%%%%%%%%%%%%%%%%%%%%%%%%%%%%%%%%%%%%%%%%%%%%%%%%%%%%%%%%%%%

%%%%%%%%%%%%%%%%%%%%%%%%%%%%%%%%%%%%%%%%%%%%%%%%%%%%%%%%%%%%%%%%%%%%%%%%%

\section{Symmetric form of the SYM parton interaction}
\setcounter{equation}{0}

In our application we have $\ell = 1$, the corresponding labels
referring to the conformal weight will be omitted in the following.

The square bracket in first term in (\ref{vertexsym})
can be written as $J_T$ (\ref{JT}). Indeed, after applying (\ref{Jf})
and adding the disconnected contributions
allowing to write this expression in terms of $\tilde J_{11\p}^{(1)} $ and
$\tilde J_{22\p}^{(1)} $ we have
\be
\D_{1\p} D_1 D_{2\p} \theta_{11\p} \theta_{22\p} \tilde J_{11\p}^{(1)}
= - D_{1\p} D_{2\p} \{ D_1 D_{1\p} \theta_{11\p} \theta_{22\p}
\tilde J_{11\p}^{(1)} \} =
 D_{1\p} D_{2\p}  \D_{1\p} J_{T1}
\ee
by comparing the results of the differentiations $D_1 D_{1\p} $
in the braces with
$\D_{1\p} J_{T1}  $. We can check also the coincidence of corresponding
coefficients in the expansion with respect to the odd coordinates.
The one at $\theta_1 \theta_2$ is proportional to $J_{11\p}^{(2)} +
J_{22\p}^{(2)}$ and the one at
$\theta_{1\p} \theta_{2\p}$ is proportional to $J_{11\p}^{(1)} +
J_{22\p}^{(1)}$.

By applying the same transformation to the second square bracket in
(\ref{vertexsym}) it becomes
\bea
D_{1\p} D_{2\p} \ \ J_T + {(\D_1 D_1 - \D_{1\p} D_{1\p} )
(\D_2 D_2 - \D_{2\p} D_{2\p} ) \over (\D_1 + \D_{1\p} )^2}
\theta_{11\p} \theta_{22\p} \tilde J_{111} \cr -
(D_1 + D_{1\p}) (D_2 + D_{2\p}) {\D_1 \D_2 \over (D_1 + \D_2 )^2 }
\theta_{11\p} \theta_{22\p} \tilde J_{0}
\eea

By the crossing relation (\ref{crossing}) we write the last term
in (\ref{vertexsym}) with  $({\cal A}^*_1 T^a {\cal A}_{2} ) $ $
({\cal A}_{1\p} T^a {\cal A}^*_{2\p} ) $ as two terms to be added to the
one with $({\cal A}^*_1 T^a {\cal A}_{1\p} ) $ $
({\cal A}_{2} T^a {\cal A}^*_{2\p} ) $ and
 $({\cal A}^*_1 T^a {\cal A}^*_{1\p} ) $ $  ({\cal A}_{2} T^a {\cal
A}_{2\p} ) $. As the result (\ref{vertexsym}) reads
\bea
\label{vertexs2}
J_T
 \{ ({\cal A}_1^* T^{a} D {\cal A}_{1^{\prime}} )\
  ({\cal A}_2^* T^{a} D {\cal A}_{2^{\prime}} ) +
({\cal A}_1^* T^{a} D {\cal A}_{1^{\prime }} ) \ ({\cal A}_2 T^{a}
D {\cal A}_{2^{\prime }}^* ) \}
\eea
$$
+ \{ {(\D_1 D_1 - \D_{1\p} D_{1\p} )
(\D_2 D_2 - \D_{2\p} D_{2\p} ) \over (\D_1 + \D_{1\p} )^2}
\theta_{11\p} \theta_{22\p} \tilde J_{111} - $$
$$
(D_1 + D_{1\p}) (D_2 + D_{2\p}) {\D_1 \D_2 \over (D_1 + \D_2 )^2 }
\theta_{11\p} \theta_{22\p} \tilde J_{0}  $$
$$
+ (D_1 D_{1\p} + D_2 D_{2\p} ) {\D_1 \D_2 \over (D_1 + \D_2)^2 }
\theta_{12} \theta_{1\p2\p} \tilde J_{0}
\} \ \ \
({\cal A}_1^* T^{a} D{\cal A}_{1^{\prime }} ) \ ({\cal A}_2 T^{a}
D{\cal A}_{2^{\prime }}^* )
$$
$$
 - \{ {(D_1 + D_{1^{\prime}} ) ( D_2 + D_{2\p} ) } \  \theta_{11\p}
\theta_{22\p} \tilde J_{111} + $$
$$
 (D_1 D_{2\p} + D_2 D_{1\p} ) {\D_1 \D_2 \over (D_1 + \D_2)^2 }
\theta_{12} \theta_{1\p2\p} \tilde J_{0}
\}
\ \ ({\cal A}_1^* T^{a} D{\cal A}_{1^{\prime }}^* ) \ ({\cal A}_2 T^{a}
D {\cal A}_{2^{\prime }} )
$$

Now we look for  expressions of the remaining terms in (\ref{vertexs2}),
besides of the ones going with $J_T$, in terms of  conformal symmetric
kernels.
We try in particular the kernels with weight $\ell = 1$ and $d =
- \frac{1}{2} $ defining
\be
\label{JVA}
J_{V+A} = (1 + 2 \ r_{\theta t})  \ \tilde K^{(-\frac{1}{2}, 0)}
\ee
and the kernel at the lower limit of the regularity region
$(d \rightarrow - \frac{3}{2} $)
$J_V$ (\ref{JV}) because in their odd coordinate decomposition
the kernels of the QCD result (\ref{hgg},  \ref{hff}) appear.

Indeed we observe that the following relations hold
\bea
 (\D_1 D_1 - \D_{1\p} D_{1\p} )
(\D_2 D_2 - \D_{2\p} D_{2\p} )
\theta_{11\p} \theta_{22\p} \tilde J_{111} - \cr
(D_1 + D_{1\p}) (D_2 + D_{2\p}) \D_1 \D_2
\theta_{11\p} \theta_{22\p} \tilde J_{0} \cr
+ (D_1 D_{1\p} + D_2 D_{2\p} ) (\D_1 + \D_{1\p} )^2
{\D_1 \D_2 \over (D_1 + \D_2)^2 }
\theta_{12} \theta_{1\p2\p} \tilde J_{0}
\ = \ \cr
 - (\D_1 + \D_{1\p} )^2  D_{1\p} D_{2\p}
(J_V + 2 J_{V+A}), \cr
{(D_1 + D_{1^{\prime}} ) ( D_2 + D_{2\p} ) } \  \theta_{11\p}
\theta_{22\p} \tilde J_{111} + \cr
 (D_1 D_{2\p} + D_2 D_{1\p} ) {\D_1 \D_2 \over (\D_1 + \D_2)^2 }
\theta_{12} \theta_{1\p2\p} \tilde J_{0}
\ = \ \
 D_{1\p} D_{2\p}  J_V
\eea
As the result we obtain the manifest superconformal expression
 of the SYM effective vertices in the Bjorken limit
\bea
\label{hsusy}
\int
[J_T + 2 w_f \theta_{11\p} \theta_{22\p} \ \delta(z_{11\p})
\ \delta(z_{22\p}) ]  \ \ \ \cr
\{ ({\cal A}_1^* T^{a} D {\cal A}_{1^{\prime}} )\
  ({\cal A}_2^* T^{a} D{\cal A}_{2^{\prime}} ) +
({\cal A}_1^* T^{a} D{\cal A}_{1^{\prime }} ) \ ({\cal A}_2 T^{a}
D {\cal A}_{2^{\prime }}^* ) \} \cr
- [J_V + 2 J_{V+A} ] \ 
({\cal A}_1^* T^{a} D{\cal A}_{1^{\prime }} ) \ ({\cal A}_2 T^{a}
D{\cal A}_{2^{\prime }}^* )  \ 
- J_V \  ({\cal A}_1^* T^{a} D {\cal A}^*_{1^{\prime }} ) \
({\cal A}_2 T^{a} D {\cal A}_{2^{\prime }} )
\eea

The integation is over the super coordinates ($z_i, \theta_i$) of the
points $i= 1,2,1\p, 2\p $. The dependence of the super fields
${\cal A}_i$ on the points is abbreviated by the corresponding subscript.
The notation of the brackets with the generators $T^a$ is as in
(\ref{brackets}). The superconformal kernels $J_T, J_V, J_{V+A}$
depend on the coordinates of the 4 points and have been defined in
(\ref{JT}, \ref{JV}, \ref{JVA}).

The corresponding result of \cite{Belitsky:2004yg} concerns the first term in (\ref{hsusy})
proportional to $J_T$,
 the remaining terms cannot be obtained as a reduction of the one of
${\cal N } = 4 $ SYM.

Supermultiplet  parton states are defined in straightforward analogy to
the two-gluon parton states, e.g.
\be
\label{statess}
| \bar n, \nu, m; +,- > \ = \ \int dz_1 dz_2 d\theta_1 d\theta_2
\psi_{\bar n, \nu}^{(m)} (z_1, \theta_1, z_2, \theta_2 ) \
\D^{-1} {\cal A}_1^*   \D^{-1} {\cal A}_2
\ |0>
\ee
Here it is enough to label the helicity states (by $\pm $) besides of the
quantum numbers of the superconformal representation $\bar n, \nu; \
(\bar n = 0, 1, ..., \nu = 0,1 )$ and the representation level $m = 0,
\frac{1}{2}, 1, ...$. Eigenstates of the effective vertex operator
(\ref{hsusy}) are $
| \bar n, \nu, m; +,- > \pm | \bar n, \nu, m; +,- >,
| \bar n, \nu, m; +,+ > | \bar n, \nu, m; -,- > $.
The eigenvalues can be obtained by (\ref{eigenv}).

The relation between the super multiplet states (\ref{statess}) and the
parton states
defined in (\ref{pstates}) is obtained by doing the $\theta$ integrations in
(\ref{statess}). In particular,
$|\bar n, 0; f^* f>, | \bar n-1, 0; A^*,A>,
|\bar n, 0; f^* A>, | \bar n, 0; A^*, f> $
are linear combinations of the 4 supermultiplet states
$|\bar n,0, 0; +,- >,$ $ |\bar n,1 ,0 ; +,- >, $ $
|\bar n,0,\frac{1}{2} ; +,- >,$ $|\bar n -1,1 ,\frac{1}{2} ; +,- > $.
i.e. 3 adjacent superconformal representations contribute to the
conformal parton states of a given weight and different parton type
(gluon, fermion). For this example
we have to substitute for the lowest states
$\psi_{\bar n, \nu}^{(0)}$ from (\ref{lowest2}) and
for the next-to-lowest states  $\psi_{\bar n, \nu}^{(\frac{1}{2})}$
from  (\ref{psi12}) to obtain
\bea
|\bar n, 0;0;+,-> = \bar n |\bar n -1, 0; A^* A> + c^2 |\bar n ,0;f^* f>,
\cr
|\bar n, 1;0;+,-> = -c ( |\bar n , 0; A^* f> +  |\bar n ,0;f^* A> ),
\cr
|\bar n, 0;\frac{1}{2};+,-> = -c (\bar n +2)
( |\bar n, 0; A^* f> -  |\bar n ,0;f^* A> ),  \cr
|\bar n-1, 1;\frac{1}{2} ;+,-> = (\bar n +3) |\bar n -1, 0; A^* A>
- c^2 |\bar n ,0;f^* f>,
\eea
The result implies analogous relations between the super multiplet
eigenstates and eigenvalues of (\ref{hsusy})
and the eigenstates and eigenvalues of the particular
gluon and fermion vertices, including the mixed gluon-fermion and
annihilation-type interaction kernels, which can be written as two
$2 \times 2$ matrix relations established already in \cite{BuFKL}.

\section{Summary}

Usually the renormalization of composite operators and the DGLAP/ERBL
evolution of (generalized) parton distributions is
formulated in terms of a basis of operators with their anomalous
dimensions or as a set of evolution kernels in longitudinal momenta.
Here we formulate instead quartic vertices of parton fields involving
kernels in positions on the light ray, Fourier conjugate to the
longitudinal momenta. This formulation allows to represent  the (super)
conformal symmetry of the one-loop parton interaction in the most
explicit way. The kernels are (super) conformal 4-point functions.
The effective parton interaction vertices provide a compact formulation
of the leading twist exchange, where the twist equals to the number of
exchanged partons.

This  manifest symmetric representation  can be obtained as a
transformation of the conventional ones, e.g. by  Fourier
transformation of the momentum kernels. However it is not
straightforward to recover the symmetric light-ray kernels in this way.
The calculation following the effective action concept, starting from
light cone form of the action, leads to the symmetric (one-loop) 
result via two
standard loop integrals by a few transformations, combinig terms to
remove spurious infrared divergencies and writing the vertices in terms
of the conformal primary parton fields.

In the present paper we have applied this scheme \cite{DKBj} to $ {\cal N}= 1 $
super Yang-Mills theory emphasizing the close analogy to the gluodynamic
(${\cal N} = 0 $) case. The light cone action has been written in a
simple super field form involving one odd ($\theta $) variable only. The
calculation has been done in a form symmetric with respect to super
translations (acting on $\theta$ and the light ray coordinate). The
superconformal symmetry of the effective parton interaction has been
formulated in the framework of the symmetry algebra $osp(2|1)$.

Parton super multiplet states are written in terms of wave functions on
the super extended light ray $(z, \theta)$. They  decompose into a
polynomial basis corresponding to irreducible lowest weight
representations of the symmetry algebra. There is a duality of the multi
parton wave functions  to composite operators of the parton fields by a
symmetric scalar product.

The generic form of superconformal 4-point functions, being the
kernels of symmetric two-parton operators, provides the guideline for
identifying the wanted symmetric vertices.
We find it convenient to start from symmetric closed-contour integral
operators with kernels written in analytic power-like terms and to
consider the kernels of integral operators on the light-ray as the
result of the former by contour contraction.

Finally the effective  vertices for ${\cal N } = 1 $
super Yang Mills leading
parton interaction are written in 3 terms involving particular 
superconformal kernels and contributing differently to the 3 different
helicity states of the interacting parton pair.

\section*{Acknowledgements}

The author is grateful to
S. Derkachov,  D. Karakhanyan and L.N. Lipatov for useful discussions.

\end{document}